\documentclass[12pt]{iopart}
\usepackage{graphics}

\begin{document}

\title[Glass transition of confined fluids: Mode-coupling
theory]{Liquid-glass transition of confined fluids: Some insights from
a mode-coupling theory}

\author{V Krakoviack}

\address{Laboratoire de Chimie, UMR CNRS 5182, \'Ecole Normale
Sup\'erieure de Lyon, 46, All\'ee d'Italie, 69364 Lyon Cedex 07,
France}

\ead{vincent.krakoviack@ens-lyon.fr}

\begin{abstract}
The dynamics of confined glassforming liquids is discussed on the
basis of the recent extension of the mode coupling theory for the
liquid-glass transition to the model of the quenched-annealed binary
mixture. It is in particular shown that, in confinement, the density
correlation functions always decay to a non-zero infinite time value,
even in the fluid state, and some clarification is given about the
question of the relation between structure and dynamics in confined
fluids.
\end{abstract}


\nosections

In the past few years, there has been a strong interest for the
dynamics of liquids under confinement and more specifically for their
structural glass transition, which has been investigated by a variety
of experimental techniques and by computer simulation
\cite{proceedings,review05}. One of the main goals of these studies
was to improve our understanding of the concept of cooperativity, a
key ingredient of many glass transition theories
\cite{heterogeneities}. Indeed, there are now many evidences that, in
the bulk, the dynamics of deeply supercooled liquids is strongly
inhomogeneous and that correlated clusters of molecules, the so-called
dynamical heterogeneities, play a crucial role in the slowing-down of
the dynamics when the temperature is decreased. But, up to now, many
aspects of the characterization of these dynamical heterogeneities
have remained elusive. For instance, there is no clear consensus on
their shape, their size and their evolutions with temperature.

This is where the interest for confined glassforming systems came
in. Indeed, confinement is a means to introduce geometrical
constraints and new characteristic lengthscales (pore size, film
thickness\dots) in the system under study. Thus, by looking at the way
the dynamics is modified under confinement compared to the bulk, one
can hope to gain some insight on the properties of the dynamical
heterogeneities. For instance, in the simplest scenarios, one expects
from finite size effects a cutoff on the slowing-down of the dynamics
as temperature is varied, when the typical size of the heterogeneities
in the bulk would become larger than the characteristic lengthscale of
the confining medium. Confinement would thus provide an indirect probe
of the properties of the dynamical heterogeneities.

It turns out that the situation is more complex. Indeed, the previous
line of reasoning requires that the physical phenomena which are
specific of confined systems have a negligible impact on the dynamics
of the imbibed fluid or at least that their influence is sufficiently
well known that it can subtracted from the results.  This is usually
not the case and strong confinement effects are observed, in
particular at the fluid-solid interface where structured layers of
almost immobile molecules are often formed. So this is in fact the
problem of dynamics under confinement as a whole which has to be
addressed and not only at the level of a simple modification of the
bulk dynamics.

Theories dealing with the dynamics in confinement are far less
developed than for the bulk and they usually are of a phenomenological
nature \cite{review05}. There is yet a strong need for elaborate
microscopic theories in this field.  Indeed, the variety of systems to
consider is immense. Porous media can differ in the shape, size, size
distribution, connectivity, etc, of their pores. They can be made of
various materials, leading to a wide range of fluid-solid interactions
which adds to the already great variability of intermolecular
interactions met with usual glassformers. A reasonable theoretical
approach, able to catch many of these subtleties, would thus be of a
great help. Applied to various models, it would allow to explore
thoroughly the phenomenology of confined glassforming systems and
maybe to disentangle the different physical effects which interplay in
these systems.

Recently, a step in this direction has been made with the extension of
the mode coupling theory (MCT) for the liquid-glass transition
\cite{bengotsjo84jpc,leshouches,gotsjo92rpp,got99jpcm} to a
particular class of confined systems, the so-called
``quenched-annealed'' (QA) binary mixtures \cite{kra05prl}. In these
systems, first introduced by Madden and Glandt \cite{MG1988}, the
fluid molecules equilibrate in a matrix of particles frozen in a
disordered configuration sampled from a given probability
distribution.  This class of models, which describe situations of
statistically homogeneous and isotropic confinement, is thought to be
able to reproduce most of the physics of fluids confined in materials
like Vycor, controlled porous glasses or aerogels, and, in fact, the
fluid dynamics and glass transition in some of its instances have been
the subject of recent studies by molecular dynamics simulations
\cite{galpelrov02el,kim03el,chajagyet04pre}.

In this paper, we will give a short presentation of the proposed
extension of the MCT to QA systems (a more detailed account is given
in reference \cite{kra05prl}) and, on the basis of this approach, we
will discuss a few aspects of the physics of confined fluids which are
of relevance for the interpretation of experimental and computer
simulation results.

Before dealing with dynamics, one first has to consider some
peculiarities of the statics of QA systems. QA mixtures are systems
with quenched disorder, so that their theoretical description requires
two types of averages, a thermal average denoted by $\langle \cdots
\rangle$, taken for a given realization of the matrix, and a disorder
average over the matrix realizations, denoted by $\overline{\cdots}$,
to be taken after the thermal average.  Like in the bulk, one is
interested in the Fourier components of the microscopic fluid density,
or, in short, density fluctuations, defined as
$\rho^f_\mathbf{q}(t)=\sum_{j=1}^{N_f} e^{i \mathbf{q}
\mathbf{r}_j(t)}$, where $\mathbf{q}$ denotes the wavevector, $N_f$ is
the fluid particle number and $\mathbf{r}_j(t)$ is the position of
fluid particle $j$ at time $t$. A significant difference with the bulk
is that, for a given matrix realization, the translational invariance
of the system is broken by the presence of the quenched
component. This results in non-zero average density fluctuations at
equilibrium, i.e., $\langle \rho^f_\mathbf{q} \rangle \neq 0$. It is
only after the disorder average that the symmetry is restored, leading
to $\overline{\langle \rho^f_\mathbf{q} \rangle} = 0$, hence the
description of the model as \emph{statistically} homogeneous and
isotropic. This property has a well known impact on the equations
describing the structural correlations in such systems, for instance
the so-called replica Ornstein-Zernike (OZ) equations
\cite{G1992,RTS1994}, where it leads to the splitting of the total and
direct correlation functions of the fluid, $h^{ff}(r)$ and
$c^{ff}(r)$, respectively, into two contributions, connected
[$h^{c}(r)$ and $c^{c}(r)$] and blocked or disconnected [$h^{b}(r)$
and $c^{b}(r)$].  The separation of $h^{ff}(r)$ into two terms leads
to a similar property of the fluid structure factor $S^{ff}_q=
\overline{\langle \rho^f_\mathbf{q} \rho^{f}_\mathbf{-q} \rangle} /
N_f =1+n_f \hat{h}^{ff}_q$, where $n_f$ is the fluid number density,
which can be expressed as $S^{ff}_q = S^{c}_q + S^{b}_q$ with $S^{c}_q
= \overline{\langle( \rho^f_\mathbf{q} -
\langle\rho^f_\mathbf{q}\rangle)
(\rho^{f}_\mathbf{-q}-\langle\rho^{f}_\mathbf{-q} \rangle)\rangle} /
N_f=1+n_f \hat{h}^{c}_q$ and $S^{b}_q =
\overline{\langle\rho^f_\mathbf{q}\rangle \langle\rho^{f}_\mathbf{-q}
\rangle}/N_f=n_f \hat{h}^{b}_q$, where $\hat{f}_q$ denotes the Fourier
transform of $f(r)$.  To fix all the notations, we define here the
matrix-matrix and fluid-matrix structure factors and total correlation
functions as well, which are given by $S^{mm}_q = \overline{\langle
\rho^m_\mathbf{q} \rho^{m}_\mathbf{-q}\rangle}/ N_m=1+n_m
\hat{h}^{mm}_q$ and $S^{fm}_q = \overline{\langle \rho^f_\mathbf{q}
\rho^{m}_\mathbf{-q} \rangle} / \sqrt{N_f N_m}= \sqrt{n_f n_m}
\hat{h}^{fm}_q$, where $N_m$ is the matrix particle number, $n_m$ is
the matrix number density, and $\rho^m_\mathbf{q}=\sum_{j=1}^{N_m}
e^{i \mathbf{q} \mathbf{s}_j}$, where $\mathbf{s}_j$ is the fixed
position of matrix particle $j$, is the $\mathbf{q}$ Fourier component
of the quenched microscopic matrix density.

The identification of two types of static fluid correlations has
significant implications for the dynamics. Indeed, if one forgets for
a moment the possibility of a dynamical ergodicity breaking, one
expects, using standard arguments, that $\lim_{t\to\infty}
\overline{\langle \rho^f_\mathbf{q}(t) \rho^{f}_\mathbf{-q} \rangle} =
\overline{\langle\rho^f_\mathbf{q}\rangle \langle\rho^{f}_\mathbf{-q}
\rangle }$, i.e., the normalized total density fluctuation
autocorrelation function $\phi^T_q(t)= \overline{\langle
\rho^f_\mathbf{q}(t) \rho^f_\mathbf{-q}\rangle}/ (N_f S^{ff}_q)$ does
not decay to zero at long times, but rather
\begin{equation*}
\lim_{t\to\infty} \phi^T_q(t)= \frac{S^{b}_q}{S^{ff}_q}.
\end{equation*}
This general result only derives from the fact that the fluid is
plunged in an inhomogeneous external potential, which here is due to
the matrix. The existence of a non-zero limit thus is independent of
the model and in particular it does not depend on the fluid-matrix
interaction. To use the language of scattering experiments, it means
that one should always expect to measure non-vanishing intermediate
scattering functions or that an elastic contribution has to be present
in the dynamical structure factors \cite{zorn02,alba03}. We stress
here that this is a true static phenomenon, and not one related to
some ultra slow dynamical process.

We can now turn to the dynamical theory. Following the preceding
discussion, the proper dynamical variable to consider is the relaxing
part of the fluid density fluctuations $\delta\rho^f_\mathbf{q}(t) =
\rho^f_\mathbf{q}(t)-\langle \rho^f_\mathbf{q} \rangle$, rather than
the full $\rho^f_\mathbf{q}(t)$. Then, using standard projection
operator methods \cite{leshouches}, the equations of the MCT for QA
systems are obtained \cite{kra05prl}. They are equations for the time
evolution of the normalized connected density fluctuation
autocorrelation function, $\phi_q(t) = \overline{\langle \delta
\rho^f_\mathbf{q}(t) \delta\rho^f_\mathbf{-q}\rangle}/ (N_f S^{c}_q)$,
and they consist of a generalized Langevin equation, which is the same
as for the bulk,
\begin{equation*}\label{langcoll}
\ddot{\phi}_{q}+\Omega_{q}^2 \phi_{q}+ \Omega_{q}^2 \int_0^t d\tau
M_q(t-\tau) \dot{\phi}_{q}(\tau)=0,
\end{equation*}
with $\Omega_{q}^2=q^2 k_B T /(m S^{c}_q)$, where $m$ is the mass of
the fluid particles, $T$ the temperature and $k_B$ the Boltzmann
constant, and of an expression for the memory kernel $M_q(t)=\Gamma_q
\delta(t) + M^{(\mathrm{MC})}_q(t)$, with
\begin{equation*}\label{kerncoll}
M^{(\mathrm{MC})}_q(t)= \int \frac{d^3\mathbf{k}}{(2\pi)^3} \left[
V^{(2)}_{\mathbf{q},\mathbf{k}} \phi_{k}(t) \phi_{|\mathbf{q-k}|}(t) +
V^{(1)}_{\mathbf{q},\mathbf{k}} \phi_{k}(t) \right] ,
\end{equation*}
\begin{equation*}
V^{(2)}_{\mathbf{q},\mathbf{k}} = \frac{1}{2} n_f S^{c}_q
\left[\frac{\mathbf{q}\cdot\mathbf{k}}{q^2} \hat{c}^{c}_k +
\frac{\mathbf{q}\cdot(\mathbf{q-k})}{q^2}
\hat{c}^{c}_{|\mathbf{q-k}|}\right]^2 S^{c}_k
S^{c}_{|\mathbf{q-k}|},
\end{equation*}
\begin{equation*}
V^{(1)}_{\mathbf{q},\mathbf{k}} = n_m S^{c}_q
\left[\frac{\mathbf{q}\cdot(\mathbf{q-k})}{q^2}+n_f
\frac{\mathbf{q}\cdot\mathbf{k}}{q^2} \hat{c}^{c}_k \right]^2
\frac{(\hat{h}^{fm}_{|\mathbf{q-k}|})^2}{S^{mm}_{|\mathbf{q-k}|}}
S^{c}_k.
\end{equation*}
There are strong analogies between these equations and those for the
bulk. Indeed, they share the same mathematical structure, so that most
of their properties are already known from the extensive study of
reference \cite{leshouches}. At a more detailed level, the expressions
of the vertices $V^{(2)}_{\mathbf{q},\mathbf{k}}$ and
$V^{(1)}_{\mathbf{q},\mathbf{k}}$, which determine the long time
behaviour of $\phi_q(t)$ and involve static quantities only, are very
similar to those found in the bulk, for collective and tagged particle
density relaxations, respectively \cite{bengotsjo84jpc}.

About these vertices, a specific point is worth stressing. Indeed, one
can see that the structural quantities on which they depend are not
those related to the full density fluctuations, but the connected
ones, which specifically characterize the static correlations of the
relaxing part of the density fluctuations. This result, whose
generality, we believe, goes beyond the model of QA mixtures, solves
an apparent paradox mentioned in various places and sometimes
considered as a challenge to a mode coupling description of the glass
transition in confinement, that systems with identical structural
properties can have significantly different dynamical behaviours
\cite{kim03el,schkolbin04jpcb}: In fact, only a nontrivial fraction of
the total density fluctuations, a fraction which exists on top of the
static average density fluctuations induced by the external potential
in which the fluid particles evolve, is ``active'' in the
determination of the dynamical properties of the confined fluid, and
it is the evolution of this contribution which has to be considered to
discuss the relation between the fluid structure and dynamics and not
the total structure functions of the system.

The solution of the MCT equations for QA mixtures offers no special
difficulties compared to the bulk, since we essentially deal with the
same type of equations. In reference \cite{kra05prl}, we have obtained
by the methods of reference \cite{frafucgotmaysin97pre} the dynamical
phase diagram of a simple system consisting of a fluid of hard spheres
confined in a matrix of hard spheres frozen in an equilibrium
configuration, with both types of particles having diameter $\sigma$
\cite{chajagyet04pre}. We present here typical examples of the time
evolution of the predicted density correlation functions. Like in
previous work, the system is characterized by two volume fractions
$\phi_f=\pi n_f\sigma^3/6$ and $\phi_m=\pi n_m\sigma^3/6$, and the
necessary structural quantities are computed using the Percus-Yevick
approximation \cite{G1992,MLW96JCP}. For simplicity, we neglect
$\ddot{\phi}_{q}(t)$ in the generalized Langevin equation, so that the
transient dynamics of $\phi_{q}(t)$ now has a time constant $\tau_q =
t_\mathrm{mic} S^c_q / (q \sigma)^2$ \cite{frafucgotmaysin97pre}.

\begin{figure}
\scalebox{0.85}{\includegraphics{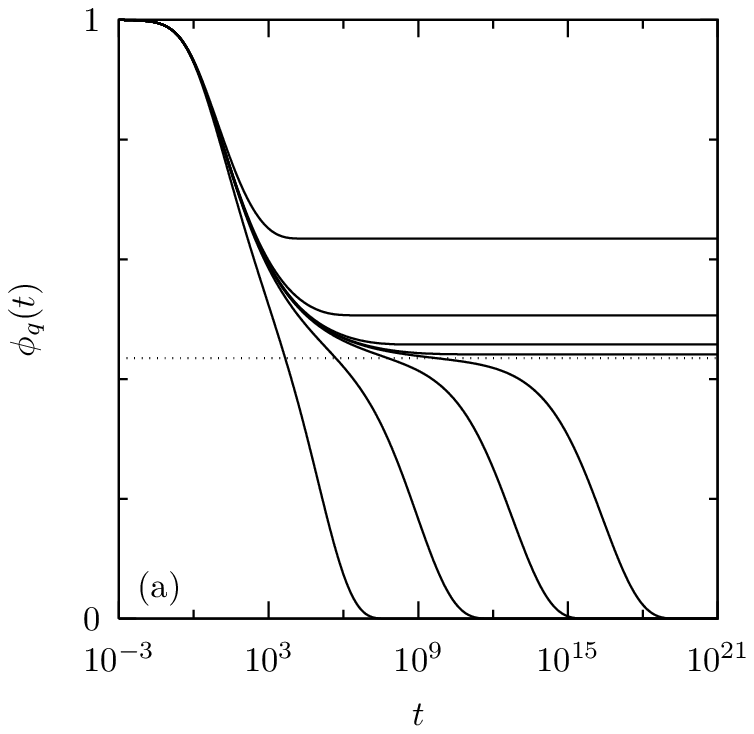}} \hfill
\scalebox{0.85}{\includegraphics{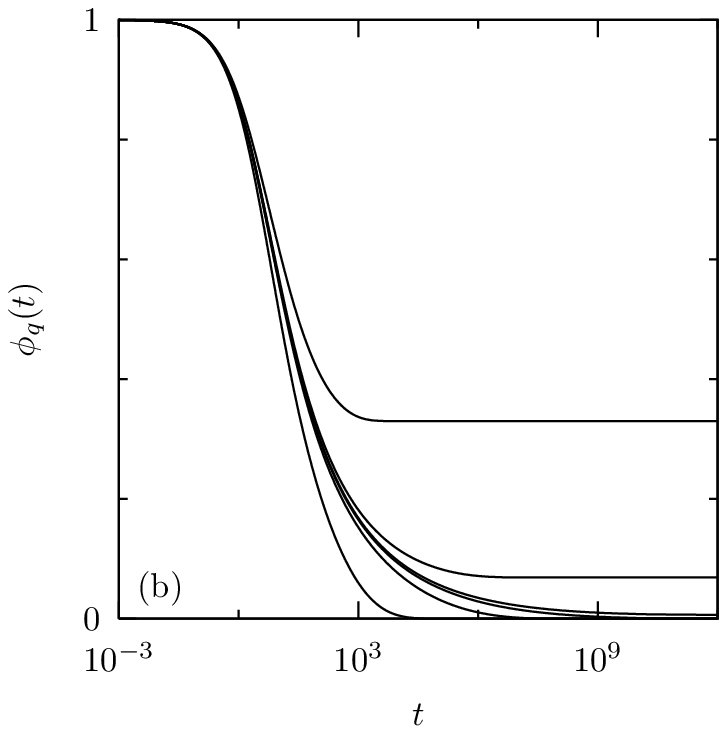}}
\caption{\label{fig1} Connected density correlation function
$\phi_{q}(t)$ for hard sphere fluids confined in matrices of identical
hard spheres frozen in an equilibrium configuration. $q\sigma\simeq
6.71$ is close to the position of the main maximum of the fluid
structure factor. We use $t_\mathrm{mic} = 160$. (a) $\phi_m=0.2$;
from bottom to top: $\phi_f = 0.99\phi_f^c$, $0.999\phi_f^c$,
$0.9999\phi_f^c$, $0.99999\phi_f^c$, $1.00001\phi_f^c$,
$1.0001\phi_f^c$, $1.001\phi_f^c$, $1.01\phi_f^c$, where
$\phi_f^c\simeq 0.672$ is the critical fluid compacity. The dotted
line shows the asymptote of $\phi_{q}(t)$ at $\phi_f^c$. (b)
$\phi_m=0.4$; from bottom to top: $\phi_f = 0.9\phi_f^c$,
$0.99\phi_f^c$, $0.999\phi_f^c$, $1.001\phi_f^c$, $1.01\phi_f^c$,
$1.1\phi_f^c$, where $\phi_f^c\simeq 0.279$ is the critical
compacity.}
\end{figure}

Two types of ergodicity breaking transitions were found in reference
\cite{kra05prl}. For small matrix densities, a type B or discontinuous
ideal glass transition line is met. It corresponds to the well known
scenario found in the bulk, showing a two-step relaxation in the
ergodic phase and a discontinuous change of the asymptotic value of
$\phi_{q}(t)$ from zero to a finite value when the transition line is
reached, corresponding to the divergence of the characteristic time of
the second relaxation step. This is illustrated in figure 1a. For
larger matrix densities, a type A or continuous ideal glass transition
line is found. Here, one observes a one-step relaxation and the
ergodicity breaking transition is continuous, i.e., the non-zero
asymptotic value of $\phi_{q}(t)$ grows continuously from zero when
the system enters in the non-ergodic domain. Such a behaviour is shown
in figure 1b.

From the knowledge of $\phi_q(t)$, one can compute the total density
correlation function $\phi^T_q(t)$, which is the quantity directly
measurable in experiments and simulations. Both functions are indeed
related by the simple formula
\begin{equation*}
\phi^T_q(t)=\frac{S^{c}_q}{S^{ff}_q} \phi_q(t) +
\frac{S^{b}_q}{S^{ff}_q}.
\end{equation*}
In figure 2 are reported the results of this transformation on the
data of figure 1. From these curves, difficulties can immediately be
anticipated if one is to compare the predictions of the mode coupling
theory with experimental or simulation data for $\phi^T_q(t)$. Indeed,
one sees that the blocked part of the static density correlations
provides a density dependent background on top of which the glassy
dynamics develops itself. It might thus be difficult to separate, in
the long time behaviour of $\phi^T_q(t)$, the evolutions which are of
purely static origin from those which characterize the glassy dynamics
of the system. This is especially critical in the case of type A
transitions, since when the system enters in the ideal glassy state
both contributions are, to first order, linear in $\phi_f-\phi_f^c$
\cite{leshouches}.

\begin{figure}
\scalebox{0.85}{\includegraphics{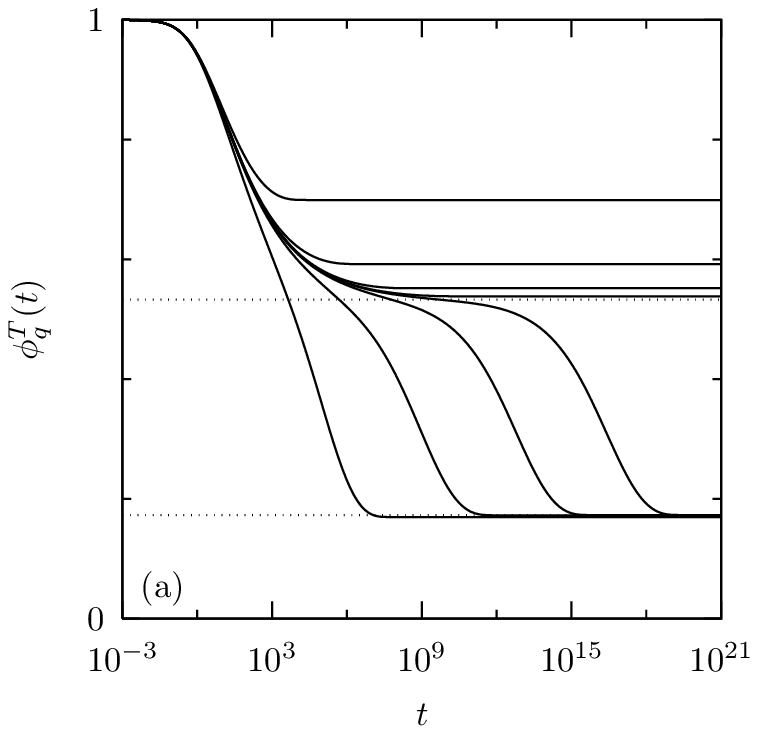}} \hfill
\scalebox{0.85}{\includegraphics{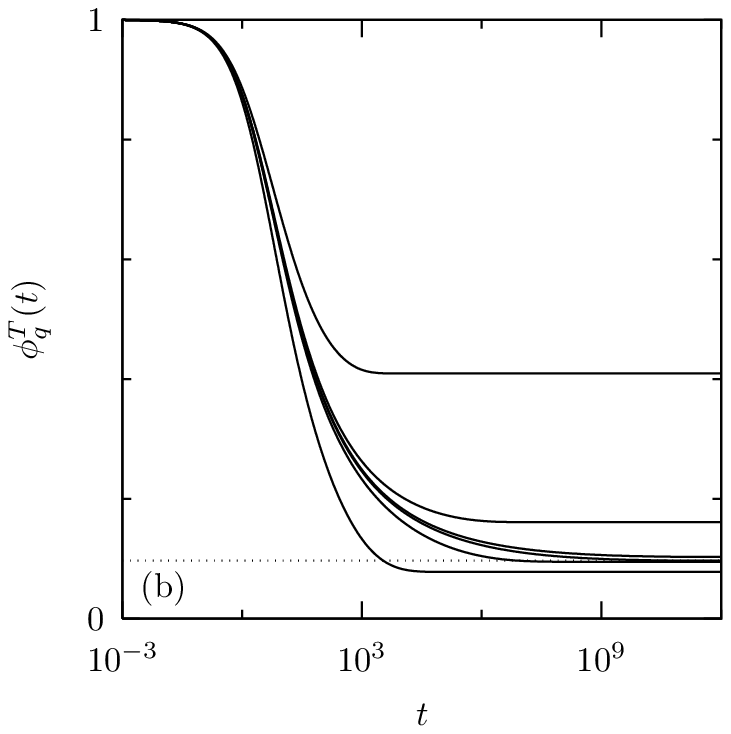}}
\caption{\label{fig2} Total density correlation function
$\phi^T_{q}(t)$ for the same systems as in figure 1. (a) $\phi_m=0.2$;
$\phi_f$'s as in figure 1a. The lowest dotted line shows the ratio
$S^{b}_q/S^{ff}_q$ at $\phi_f^c$, while the highest shows the
corresponding asymptote of $\phi^T_{q}(t)$. (b) $\phi_m=0.4$;
$\phi_f$'s as in figure 1b. The dotted line shows the ratio
$S^{b}_q/S^{ff}_q$ at $\phi_f^c$. }
\end{figure}

In conclusion, we have demonstrated how the model of the QA mixture
can be a very useful tool for a better understanding of the physics of
confined glassforming liquids. Indeed, we have shown that this model,
thanks to some crucial simplifying features, in particular the
property of statistical homogeneity, allows for a rigorous
illustration of some essential concepts which have to be taken into
account when dealing with confined systems, like the distinction
between static and relaxing density fluctuations. Now, with the
corresponding extension of the MCT which is available, it seems that
much progress could be done in our understanding of the dynamics of
confined fluid systems.


\Bibliography{99}
\bibitem{proceedings} See for instance \textit{Proceedings of the
International Workshop on Dynamics in Confinement} 2000
\textit{J. Phys. (Paris)} IV \textbf{10} Pr7-203 and
\textit{Proceedings of Second International Workshop on Dynamics in
Confinement} 2003 \textit{Eur. Phys. J.} E \textbf{12} 3-204

\bibitem{review05} For a recent review see Alcoutlabi M and McKenna G
  B 2005 \textit{J. Phys.: Condens. Matter} \textbf{17} R461

\bibitem{heterogeneities} Sillescu H 1999
\textit{J. Non-Cryst. Solids} \textbf{243} 81 \nonum Ediger M D 2000
\textit{Annu. Rev. Phys. Chem.} \textbf{51} 99 \nonum Richert R 2002
\textit{J. Phys.: Condens. Matter} \textbf{14} R703

\bibitem{bengotsjo84jpc} Bengtzelius U, G{\"o}tze W and Sj{\"o}lander
A 1984 \textit{J. Phys.} C \textbf{17} 5915

\bibitem{leshouches} G{\"o}tze W 1991 \textit{Liquids, freezing and
glass transition} edited by Hansen J-P, Levesque D and Zinn-Justin J
(Amsterdam: North Holland) p 287

\bibitem{gotsjo92rpp} G{\"o}tze W and Sj{\"o}gren L 1992
\textit{Rep. Prog. Phys.} \textbf{55} 241

\bibitem{got99jpcm} G{\"o}tze W 1999 \textit{J. Phys.:
Condens. Matter} \textbf{11} A1

\bibitem{kra05prl} Krakoviack V 2005 \textit{Phys. Rev. Lett.}
\textbf{94} 065703

\bibitem{MG1988} Madden W G and Glandt E D 1988
\textit{J. Stat. Phys.} \textbf{51} 537 \nonum Madden W G 1992
\textit{J. Chem. Phys.} \textbf{96} 5422

\bibitem{galpelrov02el} Gallo P, Pellarin R and Rovere M 2002
\textit{Europhys. Lett.} \textbf{57} 212 \nonum Gallo P, Pellarin R
and Rovere M 2003 \textit{Phys. Rev.} E \textbf{67} 041202 \nonum
Gallo P, Pellarin R and Rovere M 2003 \textit{Phys. Rev.} E
\textbf{68} 061209

\bibitem{kim03el} Kim K 2003 \textit{Europhys. Lett.} \textbf{61} 790

\bibitem{chajagyet04pre} Chang R, Jagannathan K and Yethiraj A 2004
\textit{Phys. Rev.} E \textbf{69} 051101

\bibitem{G1992} Given J A and Stell G 1992 \textit{J. Chem. Phys.}
\textbf{97} 4573 \nonum Lomba E, Given J A, Stell G, Weis J J and
Levesque D 1993 \textit{Phys. Rev.} E \textbf{48} 233 \nonum Given J A
and Stell G 1994 \textit{Physica} A \textbf{209} 495

\bibitem{RTS1994} Rosinberg M L, Tarjus G and Stell G 1994
\textit{J. Chem. Phys.} \textbf{100} 5172

\bibitem{zorn02} Zorn R, Hartmann L, Frick B, Richter D and Kremer F
2002 \textit{J. Non-Cryst. Solids} \textbf{307} 547

\bibitem{alba03} Alba-Simionesco C, Dosseh G, Dumont E, Frick B, Geil
B, Morineau D, Teboul V and Xia Y 2003 \textit{Eur. Phys. J.} E
\textbf{12} 19

\bibitem{schkolbin04jpcb} Scheidler P, Kob W and Binder K 2004
\textit{J. Phys. Chem.} B \textbf{108} 6673

\bibitem{frafucgotmaysin97pre} Franosch T, Fuchs M, G{\"o}tze W, Mayr
M R and Singh AP 1997 \textit{Phys. Rev.} E \textbf{55} 7153

\bibitem{MLW96JCP} Meroni A, Levesque D and Weis J J 1996
\textit{J. Chem. Phys.} \textbf{105} 1101 

\endbib
\end{document}